\documentclass[prl,aps,showpacs,twocolumn,floatfix]{revtex4}

\usepackage{graphicx}
\usepackage{dcolumn}
\usepackage{bm}
\usepackage{amsmath, amsthm,times}
\usepackage{epsfig}
\usepackage{amssymb}
\usepackage{txfonts}
\usepackage{subfigure}
\def\>{\rangle}
\def\<{\langle}

\begin{document}

\title{Entangled Fock States for Robust Quantum Optical Metrology, Imaging,
  and Sensing}%

\author{Sean D. Huver}
  \email{huver@phys.lsu.edu}
\author{Christoph F. Wildfeuer}
\author{Jonathan P. Dowling}
\affiliation{Hearne Institute for Theoretical Physics, Department of Physics and Astronomy, Louisiana State University,
Baton Rouge, LA 70803}
\date{\today, V1.0}

\begin{abstract}

We propose a class of path-entangled photon Fock states for robust quantum
optical metrology, imaging, and sensing in the presence of loss. We model
propagation loss with beam-splitters
and derive a reduced density matrix formalism from which we examine how photon
loss affects coherence. It is shown that particular entangled number states,
which contain a special superposition of photons in both arms of a
Mach-Zehnder interferometer, are resilient to environmental decoherence. We
demonstrate an order of magnitude greater visibility with loss, than possible
with N00N states. We also show that the effectiveness of a detection scheme
is related to super-resolution visibility.
\end{abstract}

\pacs{42.50.St, 42.50.Ar, 42.50.Dv, 42.50.-p}
\maketitle
Quantum states of light, such as squeezed states or entangled states, can be
used for metrology, image production, and object ranging, with greater
precision, resolution, and sensitivity than what is possible classically \cite{rosetta, sensor, caves,seth}. In
2000, one of the authors introduced a path-entangled number state known as the
N00N state, which is an entangled two mode state that has either all $N$
photons in one path $a$ of a Mach-Zehnder interferometer or the other
path $b$. The state may be written as
$\left|N::0\right\rangle_{_{a,b}}=(\left|N,0\right\rangle_{_{a,b}}+ \left|0,N\right\rangle_{_{a,b}})/\sqrt{2}$. With this state one can achieve super-resolution as
well as Heisenberg-limited super-sensitivity in interferometry and imaging \cite{boto, durkin}, where super-sensitivity is defined as the
ability of a particular quantum system to perform better than the
shot-noise limit, and super-resolution as performing better than the Rayleigh
diffraction limit. The super-resolution effect has been demonstrated for $N=2$
in a proof-of-principle experiment by Y. Shih in 2001 \cite{Shih}. In 2004 the group of
Steinberg demonstrated super-resolution for $N=3$, and the group of
A. Zeilinger did so for $N=4$ \cite{Bouwmeester, Walther, Mitchell}. Finally in 2007 a joint Japanese-British
collaboration demonstrated both super-resolution and sensitivity in a single
$N=4$ experiment \cite{nagata}. A large amount of publications
also investigated alternative states and detection schemes to obtain super-sensitivity and
-resolution. N00N states served for many years as a standard model for the newly
emerging fields of quantum optical metrology, imaging, and sensing. Consequently a few authors investigated the
effects of loss on the performance of quantum interferometers with N00N states. It
turns out that N00N states undergoing loss decohere very rapidly, making it
difficult to achieve super-sensitivity and resolution in an environment with
loss \cite{Parks, gilbert, rubin}.
\par In this letter we address how environmental interaction brings about
decoherence for a more generalized state with photons in both modes, and we
have discovered a class of states that improve drastically on the performance
of N00N states when loss is present. We find with these new states that while minimum sensitivity is slightly decreased, robustness against decoherence is greatly increased.
\par For practical purposes phase sensitivity is typically obtained by the linear error propagation
method, (see however Ref.~\cite{durkin}), where $\hat{O}$ represents the operator for the detection scheme being used,
\begin{equation}
\label{Eq:Phase}
\delta\phi=\frac{\Delta\hat{O}}{\left|\ \partial\langle\hat{O}\rangle / \partial\phi\right|},
\end{equation}
and $\Delta\hat{O}_{N}=\sqrt{\langle\hat{O}^{2}\rangle-\langle\hat{O}\rangle^2}$.
Eq.~(\ref{Eq:Phase}), for a N00N state with no loss, and a detection operator
$\hat{A}_{N}=\left|0,N\right\rangle\left\langle
  N,0\right|+\left|N,0\right\rangle\left\langle 0,N\right|$, which can be
implemented with coincidence measurements \cite{Mitchell}, reduces to the Heisenberg limit, $\delta\phi=1/N$, which is a $\sqrt{N}$ improvement over the shot-noise limit.
\par
\begin{figure}[h]
\includegraphics[width=7cm]{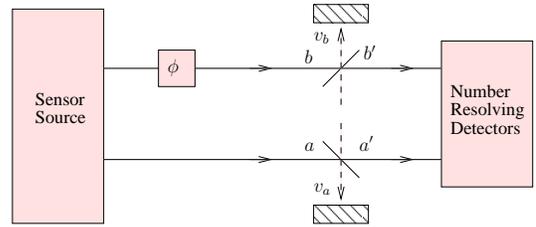} 
\caption{\label{fig:sensor} Interferometer with loss modeled by beam-splitters
  in both arms. The reflectance of the  beam-splitters determines how many
  photons one lost. An accumulated unknown phase $\phi$ is obtained due to a
  path length difference between the arms. The unitary operator for the phase
  shift is given by $\hat{U}=\mathrm{exp}(i \hat{b}^\dagger \hat{b}\phi)$. A simple
  proof shows that this operator commutes with the beam splitter
  operation. The placement of the beam splitter before the phase shift
  has been acquired therefore leads to the same result.}
\end{figure}

 
The state we now wish to examine is the following,
\begin{equation}
\label{Eq:M,M'}
\left|m :: m'\right\rangle_{_{a,b}}=\frac{1}{\sqrt{2}}\left(\left|m,m'\right\rangle_{_{a,b}}+ \left|m',m\right\rangle_{_{a,b}}\right),
\end{equation}
where we demand that $m>m'$ (we refer to this as the M\&M state). Such states
can be produced, for example, by post-selecting on the output of a pair of
optical parametric oscillators \cite{glasser}. Our setup in
Fig.~\ref{fig:sensor} is a Mach-Zehnder or an equivalent Michelson
interferometer where the details of our source and detection (such as
beam-splitters, detectors, etc.) are contained in their respective boxes. Here we are concerned primarily with how the state evolves with respect to loss, which is typically modeled by additional beam-splitters coupled to the environment \cite{shapiro}.
Similar to the approach of Ref.~\cite{rubin}, we model loss in the
interferometer with fictitious beam-splitters, but in our case these are added to both arms of the
interferometer. However we assume unit detection efficiency for the
detectors. We develop the photon statistics
as a function of beam-splitter transmittance as well as derive a reduced
density matrix, which characterizes the propagation losses inside of the interferometer. Loss is represented by
photons being reflected into the environment \cite{noise}.
The beam-splitter transforms the modes according to \cite{gerry},
\begin{eqnarray}
\hat{a}'&=&t_{a}\hat{a}+r_{a}^*\hat{a}_{v}\,,\nonumber\\ 
\hat{b}'&=&t_{b}\hat{b}+r_{b}^*\hat{b}_{v}\,,
\end{eqnarray}
where $t_u=\sqrt{T_u}\,\mathrm{exp}(i\varphi_u)$ and
$r_u=\sqrt{R_u}\,\mathrm{exp}(i\psi_u)$, $u=a,b$, are the complex transmission
and reflectance coefficients, for mode $a$ and $b$, respectively. 
The input M\&M state $|m::m'\rangle$ acquires an unknown phase shift $\phi$ and
the beam splitter transformations are applied. 
We then trace over the environmental modes, to model the photons lost, and we
obtain the reduced density matrix
$\hat{\rho}_{a',b'}=\mathrm{Tr}_{v_{a},v_{b}}\left[|\psi\rangle\langle \psi|\right]$,
which leads to
\begin{eqnarray}
\label{density}
\hat{\rho}_{a',b'}=\sum_{k=0}^{m}\sum_{l,l'=0}^{m'}|a_{k,l}|^{2}|m-k,m'-l\rangle\langle m-k,m'-l|
\nonumber\\{}+|b_{k,l}|^{2}|m'-l,m-k\rangle\langle m'-l,m-k|\nonumber\\{}
+a_{l,l'}^{*}b_{l',l}|m'-l,m-l'\rangle\langle m-l,m'-l'|\nonumber\\{}+a_{l',l}b_{l,l'}^{*}|m-l',m'-l\rangle\langle m'-l',m-l|\,.
\end{eqnarray}
Here the $a_{k,l}$ and $b_{k,l}$ coefficients are defined as
\begin{eqnarray}
\label{coefficients}
|a_{k,l}|^{2}&\equiv&\gamma_{k,l}^{2}T_{a}^{m-k}R_{a}^{k}T_{b}^{m'-l}R_{b}^{l}\,,\nonumber\\
|b_{k,l}|^{2}&\equiv&\gamma_{k,l}^{2}T_{a}^{m'-l}R_{a}^{l}T_{b}^{m-k}R_{b}^{k}\,,\nonumber\\
a_{l,l'}^{*}b_{l',l}&\equiv&\gamma_{l,l'}\gamma_{l',l}T_{a}^{\frac{m+m'-2l}{2}}R_{a}^{l}T_{b}^{\frac{m+m'-2l'}{2}}R_{b}^{l'}e^{-i\left(m-m'\right)(\phi+\varphi_b-\varphi_a)}\,,\nonumber\\
a_{l',l}b_{l,l'}^{*}&\equiv&\gamma_{l',l}\gamma_{l,l'}T_{a}^{\frac{m+m'-2l'}{2}}R_{a}^{l'}T_{b}^{\frac{m+m'-2l}{2}}R_{b}^{l}e^{i\left(m-m'\right)(\phi+\varphi_b-\varphi_a)},
\end{eqnarray} and
\begin{equation}
\label{gamma}
\gamma_{k,l}\equiv\frac{1}{\sqrt{2m!m'!}}\left(\begin{array}{c}
m\\ k
\end{array}\right)
\left(\begin{array}{c}
m'\\ l
\end{array}\right)
\left[\left(m-k\right)!k!\left(m'-l\right)!l!\right]^{1/2}.
\end{equation}
Without loss of generality we can set the transmission phases of the two beam
splitters $\varphi_a=\varphi_b=0$.
\par The reduced density matrix in Eq.~(\ref{density}) appears as an incoherent mixture plus interference terms, which survive with loss in either mode up to the limit of $m'$. The surviving interference terms all carry amplified phase information in the quantity $(m-m')\phi$. Thus the best-case minimum phase sensitivity, under no loss, is reduced from the Heisenberg limit, $\delta\phi_{N00N}=1/N$, to $\delta\phi_{m,m'}=1/(m-m')$. Although this sensitivity is less than what N00N states are capable of achieving (in the absence of loss), the fact that many more interference terms survive than with N00N states suggests that these states are more robust against photon loss.
\par To maximize phase information we choose a detection operator of the form
\begin{eqnarray}
\label{operator}
\hat{A}=\sum_{r,s=0}^{m'}|m'-r,m-s\rangle\langle m-r,m'-s| \nonumber\\+ |m-r,m'-s\rangle\langle m'-r,m-s|\,,
\end{eqnarray}
which can be implemented with number-resolving photo-detectors \cite{nam}. This operator is a more general summation over all possible cases up to $m'$ photons in either arm than the $\hat{A}_{N}$ operator (traditionally used for N00N states \cite{rosetta}).
The reduced density matrix for a N00N state is easily obtained by setting
$m=N$ and $m'=0$ in Eq.~(\ref{density}). We then obtain for the expectation value of $\hat{A}_N$ 
\begin{eqnarray}
\label{N00Nphase}
\lefteqn{\langle
\hat{A}_N\rangle=\mathrm{Tr}[\hat{A}_N\hat{\rho}_{a',b'}]=2\mathrm{Re}\left(a_{0,0}^{*}b_{0,0}\right)}\nonumber\\
&&=(T_aT_b)^\frac{N}{2}\cos(N\phi)\,.
\end{eqnarray}
The expectation value of the operator $\hat{A}$ given in Eq.~(\ref{operator}) for the M\&M state shows the benefit of having many more interference terms compared to the N00N state
\begin{eqnarray}
\label{resolution}
\langle
\hat{A}\rangle&=&\mathrm{Tr}[\hat{A}\hat{\rho}_{a',b'}]=2\mathrm{Re}\left(\sum_{l,l'=0}^{m'}a_{l,l'}^{*}b_{l',l}\right)\nonumber\\
     &=& 2 \sum_{l,l'=0}^{m'}\left|a_{l,l'}^{*}b_{l',l}\right|\cos[(m-m')\phi]\,.
\end{eqnarray}
The visibility of an attenuated mixed state in an interferometer may be
expressed as a function of the off-diagonal terms in the reduced density
matrix from Eq.~(\ref{density}) \cite{mandel},
\begin{eqnarray}
\label{mandel}
V_{\mathrm{f}}&=&2\left|\rho_{1,2}\right|=2\left|\sum_{l,l'=0}^{m'}a_{l,l'}^{*}b_{l',l}\right|=2\sum_{l,l'=0}^{m'}\left|a_{l,l'}^{*}b_{l',l}\right|\,,
\end{eqnarray}
where we call $V_{\mathrm{f}}$ the fundamental visibility and $\rho_{1,2}$ is
taken from one of the off-diagonal terms in the density matrix in
Eq.~(\ref{density}). From Eqs.~(\ref{resolution},\ref{mandel}) we see that the
expectation value of $\hat{A}$ may be written as $\langle\hat{A}\rangle=V_\mathrm{f}\cos[(m-m')\phi]$.
For a general detection operator $\hat{O}$ the amplitude of the cosine may be
smaller than $V_\mathrm{f}$, i.e.,
$V_\mathrm{f}\ge\langle\hat{O}\rangle_{\Phi=0}$. If we use a 
detection operator $\hat{O}$ so that $V_\mathrm{f}>\langle\hat{O}\rangle_{\phi=0}$, we know our detection scheme is
inefficient and we are not retrieving full phase information. We call the
visibility of a particular detection scheme the detection visibility, $V_\mathrm{det}=\langle
\hat{O}\rangle_{\phi=0}$. We see that the $\hat{A}_N$ operator, and its
more general form $\hat{A}$ in Eq.~(\ref{operator}), are both optimal for N00N and M\&M
states, respectively, and give a detection visibility equivalent to the
fundamental visibility. The fundamental visibility for a N00N state is simply
$V_\mathrm{f}=(T_aT_b)^{N/2}$, which is just the probability the N00N state
arrives at the detector with no loss.  
\par The M\&M states, with $m-m'=N$, are capable of producing the same resolution
as a $N$ photon N00N state, but at the cost of requiring $m'$ more photons to
do so, and thus they operate at a smaller shot-noise limit. As we will show,
in the presence of loss, however, many M\&M states operate below their own shot-noise limit, while N00N states of the same resolving power do not.
\par To compare a certain M\&M to a N00N state we choose the state such that
$m-m'=N$, so the amount of phase information is the same for either
state. This way our minimum phase sensitivity also starts from the same point,
$1/(m-m')=1/N.$ The true Heisenberg-limit for a M\&M state however is
determined by the total photon number in the state and is therefore given by $1/(m+m')$. The shot-noise limit for a M\&M state is $1/\sqrt{m+m'}$, while the N00N state is the usual $1/\sqrt{N}$.
\begin{figure}[h]
\includegraphics[width=6.9cm]{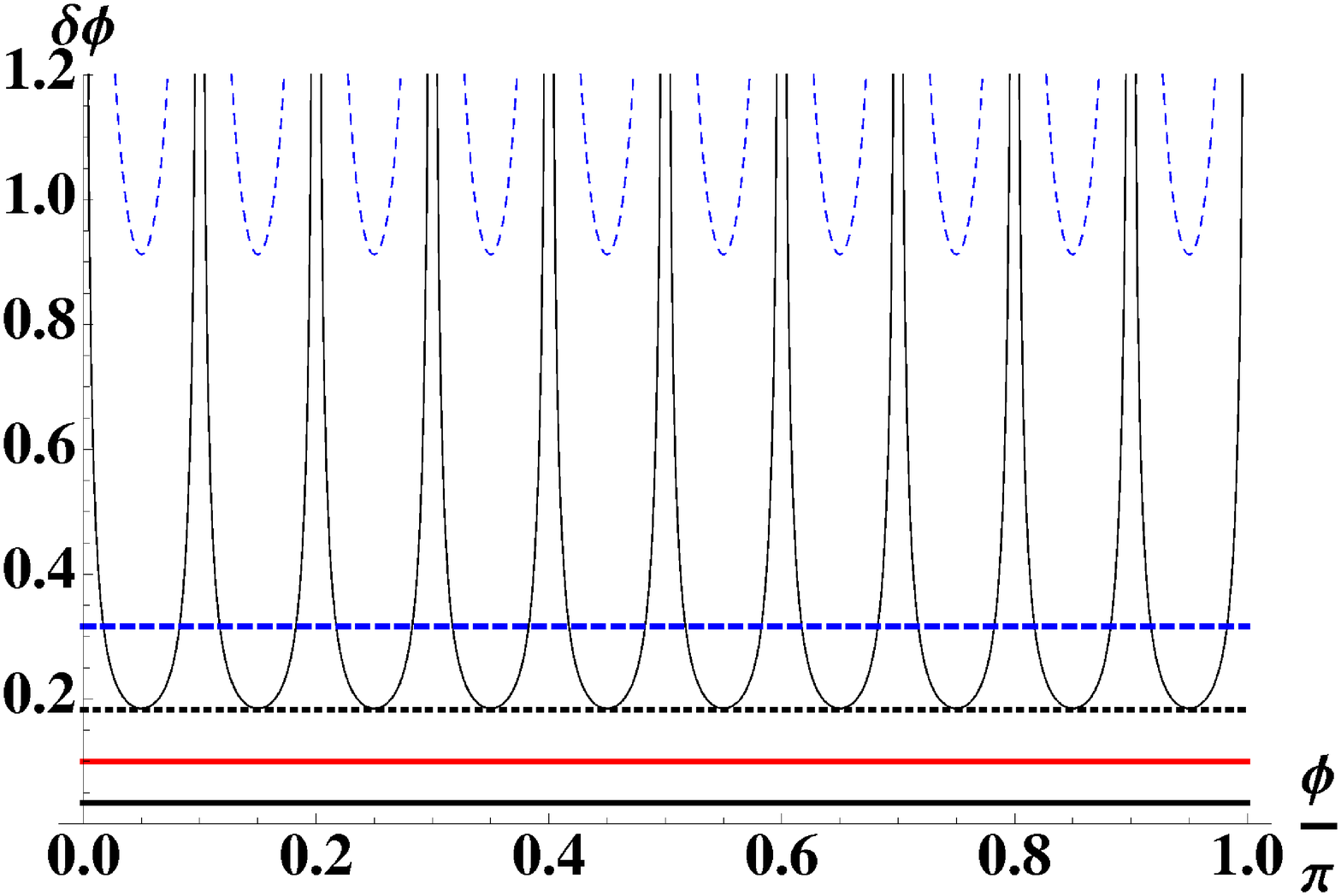}
\caption{\label{fig:phi2} Phase sensitivity $\delta\phi$ for a $\left|20::10\right\rangle$ M\&M state (curved black solid line) versus a $\left|10::0\right\rangle$ N00N state (curved blue dashed line) undergoing loss. Loss is 40\% in the long arm and zero in the delay arm. Bottom black solid line is the Heisenberg limit for $\left|20::10\right\rangle$, $1/(m+m')$. The red solid line is the Heisenberg limit for the $\left|10::0\right\rangle$ N00N state and lossless limit for $\left|20::10\right\rangle$, $1/(m-m')$. The black dotted line is the shot-noise limit for $\left|20::10\right\rangle$, while the blue dashed line is the shot-noise limit for $\left|10::0\right\rangle$. The N00N state is no longer below its shot-noise limit while the minimum phase sensitivity for the M\&M state $\left|20::10\right\rangle$ is at its respective shot-noise limit.}
\end{figure}
\par As would be the case in a practical quantum sensor, we assume loss in the long arm $b$ of the interferometer to be much greater than that of the delay arm $a$, which we assume to be under controlled loss conditions. Figure \ref{fig:phi2} is an example of a M\&M state showing more robustness to loss in phase sensitivity than an equivalent N00N state. A N00N state of $N=10$ degrades to the shot-noise limit at approximately 26\% loss in the long arm (zero loss in the delay arm), whereas a $\left|20::10\right\rangle$ M\&M state degrades to its respective shot-noise limit at larger loss, 40\% loss in the long arm (zero loss in the delay arm).
\par
\begin{figure}[h]
\includegraphics[width=6.9cm]{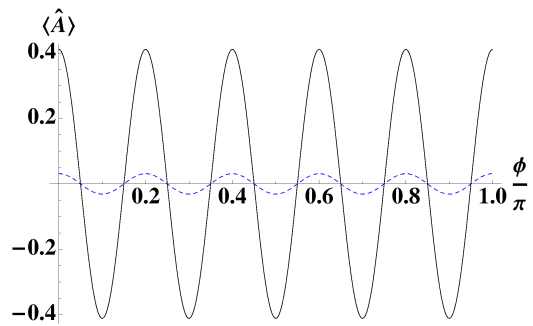}
\caption{\label{fig:resolution} Phase resolution for a $\left|20::10\right\rangle$ M\&M state (curved black solid line) versus a $\left|10::0\right\rangle$ N00N state (curved blue dashed line) undergoing 3dB of loss. Loss is 3dB in the long arm, zero in the delay arm. The amplitude of $\langle\hat{A}\rangle$, and hence the visibility of the super-resolving sub-Rayleigh fringes, for a $\left|20::10\right\rangle$ state is 41\%,  while the $\left|10::0\right\rangle$ N00N state visibility is 3.1\%.}
\end{figure}

\begin{figure}[htb]
\centering
\subfigure[]{
\includegraphics[scale=0.26]{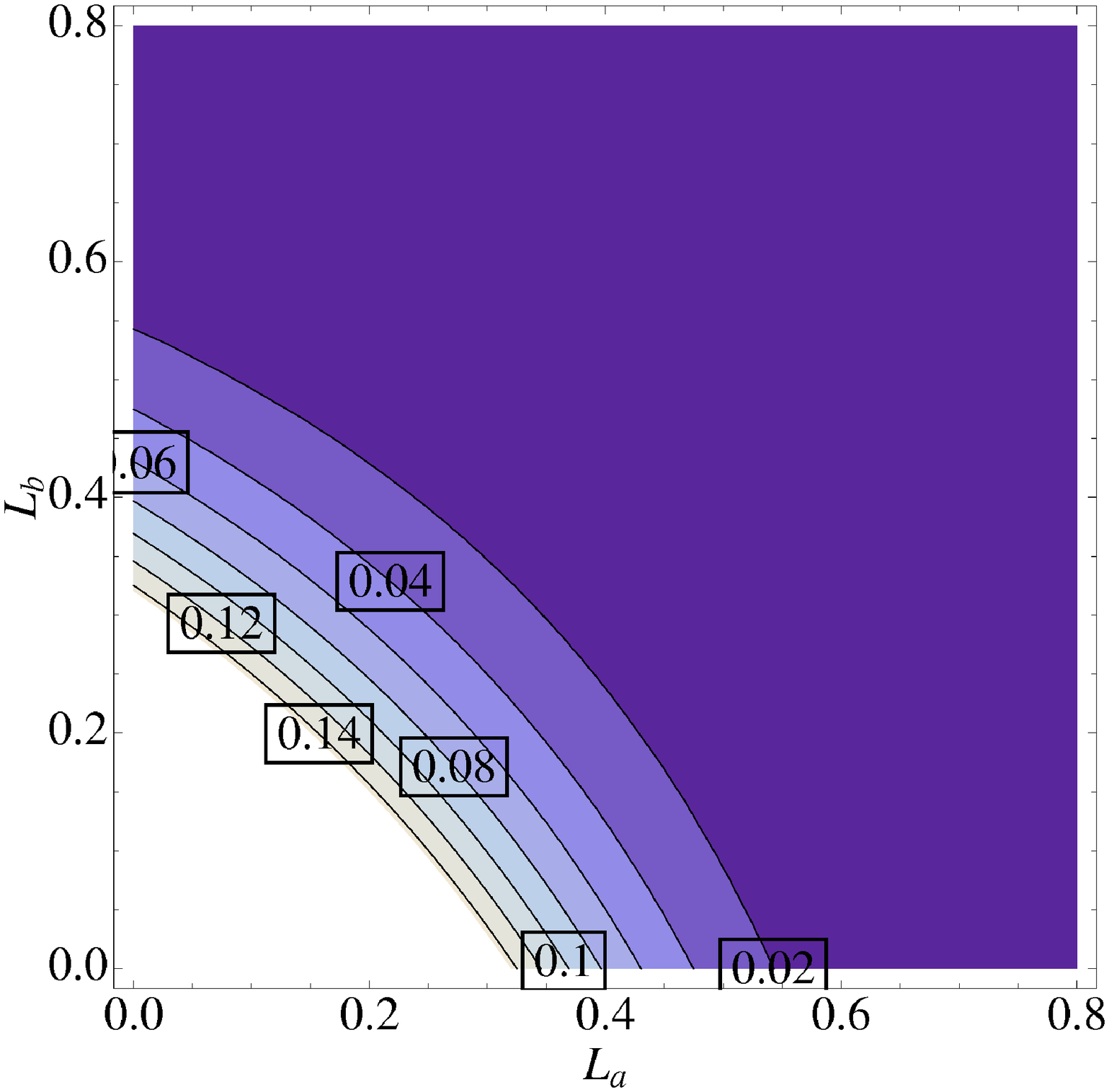}
\label{fig:vis1} }
\subfigure[]{
\includegraphics[scale=0.26]{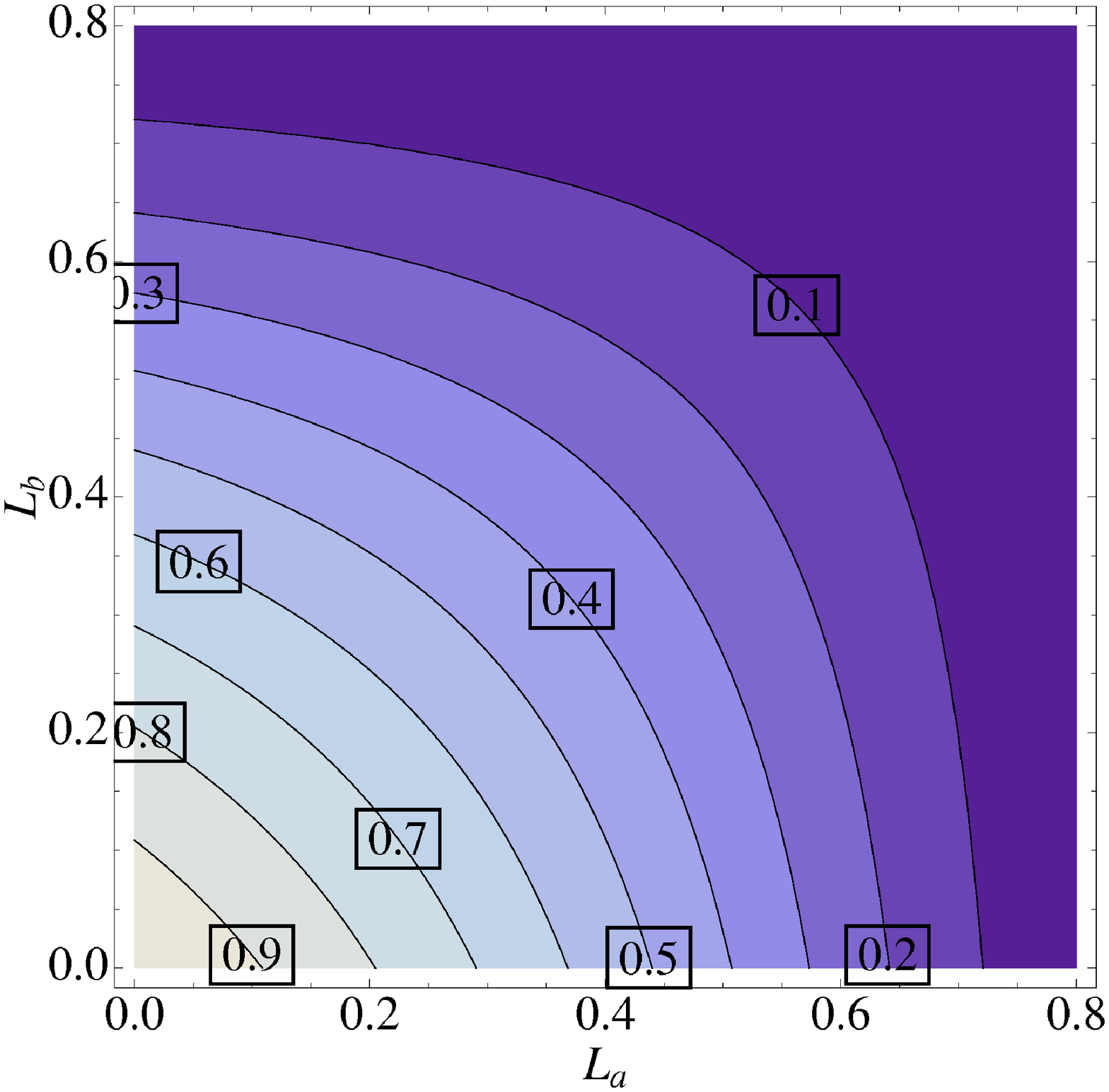}
\label{fig:vis2} }
\caption[]{Fig.~\subref{fig:vis1} Visibility for $\left|10::0\right\rangle$ as a function of
  loss in the delay and long arms, $L_{a}$ and $L_{b}$, respectively.
  Contour lines represent the value of the
  visibility, Fig.~\subref{fig:vis2} Ditto for the state $\left|20::10\right\rangle$.}
\end{figure}

 Also important is to note how $\langle\hat{A}\rangle$, and by extension,
the visibility, evolve with loss. Under lossless conditions the visibility of
a N00N or M\&M state is always one, and hence so is the amplitude of
$\langle\hat{A}\rangle$. Figure \ref{fig:resolution} shows a comparison of
$\langle\hat{A}\rangle$ for $\left|20::10\right\rangle$ and
$\left|10::0\right\rangle$ under 50\%=3dB loss in the long arm (zero in the delay arm).
We can examine the visibility as a function of loss in both arms directly with
contour plots. Figures \ref{fig:vis1} and \ref{fig:vis2} show an order of
magnitude increase in visibility for the $\left|20::10\right\rangle$ M\&M
state over the $\left|10::0\right\rangle$ N00N state. 
The improvement in visibility is greater than that seen in minimum phase sensitivity in Figure \ref{fig:phi2}. This suggests that the M\&M states are much better suited than N00N states for resolving interference fringes under loss.
\par A heuristic way to understand the improvement of M\&M states over N00N states is to consider ``which-path" information available to the environment after photon loss. For example, even a single photon lost in mode $b$ projects the N00N state from $|N,0\rangle+e^{iN\phi}|0,N\rangle$ $\longrightarrow$  $e^{iN\phi}|0,N-1\rangle$. That is, a single photon in environmental mode $b$ provides complete which-path information---the environment ``knows'' with certainty it cannot have the $|0\rangle$ component of the N00N state, which collapses the state into the separable state $|0,N-1\rangle$. In contrast, an M\&M state may lose up to $m'$ photons to the environment without complete knowledge of whether the $m$ or $m'$ component was present, and hence complete ``which-path" information is not available, and a great deal of coherence is hence preserved.
\par In comparing the M\&M states to N00N states there emerges a delicate
tradeoff in sensor performance from adding $m'$ photons to increase the number
of available output states, which contain phase information. Add too few $m'$
photons, and there will not be significant improvement. Add too many $m'$
photons, and the total number of photons required to carry the phase
information for an equivalent N00N state rises, causing the shot-noise limit
to be lowered further and reached quicker under conditions of loss (see Tab.~\ref{table:nonlin}).
\begin{table}[htb]
\caption{Comparison of visibility and minimum detectable phase for $N=10=m-m'$
  N00N state versus various M\&M states. Values are for long arm loss 3dB, and
  zero loss in delay arm. Heisenberg (HL) ($1/(m+m'))$, and shot-noise (SNL) $(1/\sqrt{m+m'})$ limits are given.}
\centering 
\begin{tabular}{|c|c|c|c|c|c|} 
\hline\hline 
$m$ & $m'$ & Visibility & $\delta\phi_{min}$ & HL & SNL\\ [0.5ex] 
\hline
10 & 0 & 3.13\% & 2.264 & 0.100 & 0.316\\ 
\hline
11 & 1 & 6.74\% & 1.051 & 0.083 & 0.289\\
\hline
12 & 2 & 10.96\% & 0.652 & 0.071 & 0.267\\
\hline
14 & 4 & 19.85\% & 0.372 & 0.056 & 0.236\\
\hline
16 & 6 & 28.11\% & 0.279 & 0.045 & 0.213\\
\hline
18 & 8 & 35.19\% & 0.238 & 0.038 & 0.196\\
\hline
20 & 10 & 41.11\% & 0.254 & 0.033 & 0.183\\
[1ex] 
\hline 
\end{tabular}
\label{table:nonlin} 
\end{table}
\par We have shown that the class of entangled Fock states with photons in both
modes, M\&M states, is more robust to loss than N00N states possessing all photons in either mode. The visibility for a M\&M state under loss may be an order of magnitude or more greater than N00N states, as well as having attenuated minimum phase sensitivities that are lower and more likely to be less than the shot-noise limit than a N00N state. While the M\&M states are not capable of reaching the Heisenberg limit of $1/N$, it seems unlikely that any state is capable of reaching this precision in the limit of practical sensing with appreciable photon loss. While M\&M states are more robust, they do appear to have loss-induced limitations as well. For many M\&M states visibility drops to approximately 10\% around the 70\% loss level in one arm, assuming perfect transmission in the other.
\par Another issue that needs consideration is how to produce M\&M states. As
of yet there is no efficient, on demand, Fock number state generator. However the output from a optical parametric amplifier (OPA) is essentially a summation of many M\&M states as well as several N00N states. We are currently analyzing the sensing capabilities for the entire output state of an OPA, as well as schemes for generating M\&M states from an OPA output with post-selection.
\begin{acknowledgments}
\par We would like to acknowledge support from the Defense Advanced Research
Projects Agency, the Army Research Office, the Intelligence Advanced Research
Projects Activity, as well as helpful discussions with H. Lee and G. A. Durkin.
\end{acknowledgments}


\begin{thebibliography}{99}
\bibitem{rosetta} H. Lee, P. Kok, and J.P. Dowling,  J. Mod. Opt., \textbf{49}, 2325 (2002).
\bibitem{caves} C.M. Caves, Phys. Rev. D, \textbf{23}, 1693 (1981).
\bibitem{sensor} K.T. Kapale, L.D. Didomenico, H. Lee, P. Kok, and J.P. Dowling, Concepts of Physics, Vol II, 225 (2005).
\bibitem{seth} V. Giovannetti, S. Lloyd, and L. Maccone, Science,
  \textbf{306}, 1330 (2004).
\bibitem{boto} A.N. Boto, P. Kok, D.S. Abrams, S.L. Braunstein,
  C.P. Williams, and J.P. Dowling, Phys. Rev. Lett., \textbf{85}, 2733 (2000).
\bibitem{durkin} G.A. Durkin and J.P. Dowling, Phys. Rev. Lett., \textbf{99}, 070801 (2007).
\bibitem{Shih} M. D'Angelo, M.V. Chekhova, and Y. Shih,
  Phys. Rev. Lett. \textbf{87}, 013602 (2001).
\bibitem{Bouwmeester} D. Bouwmeester, Nature \textbf{429}, 139 (2004).
\bibitem{Walther} P. Walther, J.-W. Pan, M. Aspelmeyer, R. Ursin,
  S. Gasparoni, A. Zeilinger, Nature \textbf{429}, 158 (2004).
\bibitem{Mitchell} M.W. Mitchell, J.S. Lundeen, and A.M. Steinberg, Nature
  \textbf{429}, 161 (2004). 
\bibitem{nagata} T. Nagata, R. Okamoto, J.L. O'Brien, K. Sasaki, S. Takeuchi, Science,
  \textbf{316}, 726 (2007).
\bibitem{gilbert} G. Gilbert, M. Hamrick, and Y.S. Weinstein, Proc. SPIE, Vol
  6573, 65730K (2007).
\bibitem{rubin} M.A. Rubin and S. Kaushik, Phys. Rev. A \textbf{75}, 053805
  (2007).
\bibitem{Parks} A.D. Parks, S.E. Spence, J.E. Troupe, and N.J. Rodecap,
  Rev. Sci. Instr. \textbf{76}, 043103 (2005).
\bibitem{glasser} R.T. Glasser, H. Cable, and J.P. Dowling, Phys. Rev. A \textbf{78}, 012339 (2008).
\bibitem{shapiro} H.P. Yuen and J.H. Shapiro, IEEE Trans. Inf. Theory, \textbf{24}, 6 (1978).
\bibitem{noise} C.W. Gardiner and P. Zoller, \emph{Quantum Noise}, (Springer, Berlin 2004).
\bibitem{gerry} C.C. Gerry and P.L. Knight, \emph{Introductory Quantum Optics}, (Cambridge University Press, Cambridge 2005).
\bibitem{nam} A.E. Lita, A.J. Miller, and S.W. Nam, Opt. Express \textbf{16}, 3032 (2008).
\bibitem{steinberg} M.W. Mitchell, J.S. Lundeen and A.M. Steinberg, Nature \textbf{429}, 161 (2004).
\bibitem{englert} P.D.D. Schwindt, P.G. Kwiat, B. Englert, Phys. Rev. A \textbf{60}, 4285 (1999).
\bibitem{mandel} L. Mandel, Opt. Lett. \textbf{16}, 1882 (1991).
\end{thebibliography}
\end{document}